\documentclass[
  aps,
  pre,
  reprint,
  superscriptaddress,
  amsmath,amssymb,
  nofootinbib,
  floatfix
]{revtex4-2}
\usepackage{booktabs}
\usepackage[labelfont=bf]{caption}
\usepackage{graphicx}
\usepackage{bm}
\usepackage{amssymb}
\usepackage{mathtools}
\usepackage{xcolor}
\usepackage{hyperref}
 \usepackage{ragged2e} 
\hypersetup{colorlinks=true,linkcolor=blue,citecolor=blue,urlcolor=blue}

\begin{document}
  \raggedbottom
\title{Susceptibility Bottlenecks in Equilibrium Propagation\\ Using Oscillator Ising Machines}

\author{{Malihe Farasat}}
\author{{Oshada Rathnayake}}
\author{{Nikhil Shukla}}
\affiliation{{Charles L. Brown Department of Electrical and Computer Engineering, University of Virginia, Charlottesville, VA 22903, USA}}

\begin{abstract}
Equilibrium propagation (EP) enables learning in physical systems by exploiting the displacement between free and weakly perturbed equilibrium states. Here, we investigate how the underlying oscillator dynamics govern the transmission of learning signals in oscillator Ising machines (OIMs). We show that the hidden-layer response can be decomposed into two distinct factors: the nudging force applied at the output and the susceptibility governing its transmission to the hidden layer. Using a bipartite OIM with interactions interpolating between phase-difference and phase-sum coupling, we investigate these factors while ensuring that all coupling forms realize the same Ising Hamiltonian at binary phase-locked states. We demonstrate that second-harmonic injection primarily influences the nudging force through phase binarization, whereas the coupling form controls the cross-layer susceptibility through the mixed curvature of the energy landscape. At equal mixing, the mixed curvature vanishes on the binary manifold, strongly suppressing hidden-layer learning signals. Numerical simulations validate these mechanisms and reveal their consequences for classification performance. Our findings establish that realizing a desired energy landscape alone is insufficient to ensure effective physical learning; the system must also support the transmission of perturbations required for learning.
\end{abstract}

    \maketitle

\section{Introduction}
Biologically inspired computing seeks to emulate the principles underlying information processing in the brain's neural circuits, where learning and adaptation emerge from local interactions and the collective dynamics of interconnected units~\cite{dillavou2022demonstration,Stern2023,momeni2025training}. This approach provides an alternative to conventional digital architectures, which typically rely on centralized computation and explicit gradient propagation. Equilibrium propagation (EP) is a biologically motivated learning framework that uses the natural relaxation dynamics of a system to generate learning signals~\cite{ScellierBengio2017}. By comparing the system's equilibrium states under free and weakly perturbed conditions, EP enables learning through the system's own dynamics, making it particularly suitable for \emph{in-situ} training of physical neural networks and oscillator-based computing systems~\cite{
Laborieux2021,laborieux2024improving,Marquardt2020}.

Coupled oscillator networks compute through the collective dynamics of interacting oscillators. This enables highly parallel processing and forms the basis of oscillator Ising machines (OIMs)~\cite{WangRoychowdhury2019, Bashar2023}. In OIMs, the phase dynamics perform gradient descent on an energy function whose coupling term encodes the Ising spin interactions. Second-harmonic injection (SHI) drives each oscillator phase toward $0$ or $\pi$, so that the network finally relaxes to a binary phase configuration representing a low-energy spin state. This natural relaxation toward equilibrium makes coupled oscillator networks a particularly attractive physical substrate for EP~\cite{fan2026hybridizing,yin2026epic,rageau2025training,wang2024training,laydevant2024training,zoppo2022equilibrium}.

One proposed approach to implementing EP using coupled oscillators, entails a complete bipartite OIM wherein one set of oscillators represents the hidden layer while the other represents the output layer. The network dynamics consist of two operational phases, free and nudge, beginning with the free phase that entails relaxation of the oscillator network to its free equilibrium (in the absence of output perturbation). In the nudge phase, a small perturbation proportional to the output error--the difference between the network output and the correct target label--is applied at the output layer, nudging the output oscillator phases closer to the target values. These target values encode the correct output for the current training example. Subsequently, the hidden layer receives its learning signal from the phase displacement of the output units between the free and perturbed equilibria. As a result, the training process depends on the perturbation that guides the network towards the learned data~\cite{ScellierBengio2017}.

However, a critical aspect of the learning effectiveness is the transmission of the perturbation to the hidden layer. This transmission depends on two factors, the force the perturbing nudge applies to the output phases, and the susceptibility of the hidden layer to this
nudging force that determines how output displacement propagates to the hidden layer. If either factor is small, the hidden layer receives little learning signal, even when the free phase forward computation is predicting the target values.

\begin{figure*}
    \centering        
               
                    \includegraphics[width=0.9\linewidth]{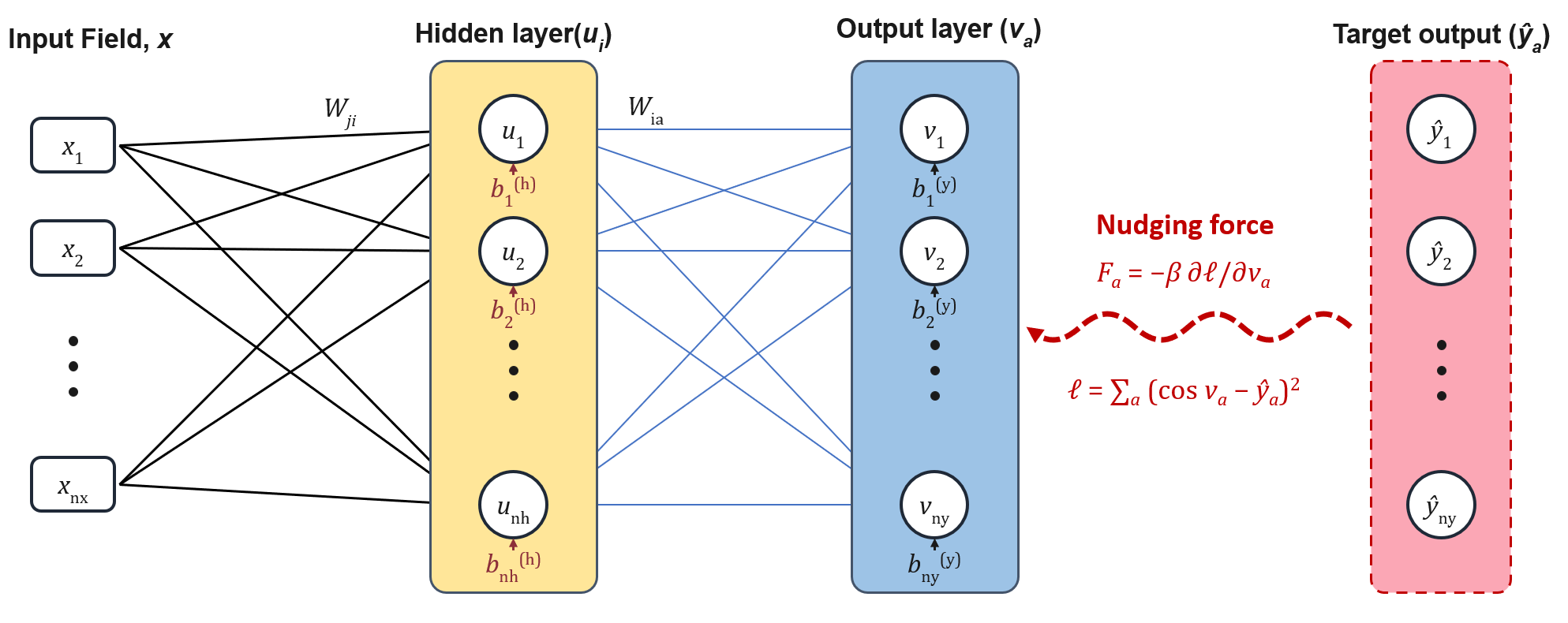}

        \caption{\justifying \small Schematic of a bipartite oscillator network trained using equilibrium propagation. The input $\bm{x}$ is held fixed and applied through the hidden-layer fields $h_i^{(h)}(\bm{x})$. The hidden and output oscillators, with phases $u_i$ and $v_a$, respectively, are coupled through the trainable weights $W_{ia}^{(hy)}$. The trainable parameters include the input-to-hidden weights $W_{ji}^{(xh)}$, hidden-to-output weights $W_{ia}^{(hy)}$, and biases $b^{(h)}$ and $b^{(y)}$. During the nudged phase, an output-dependent force drives the network toward a perturbed equilibrium.} 
    \label{fig:architecture}
\end{figure*}

Prior studies of equilibrium propagation in oscillator networks have primarily focused on establishing the feasibility of oscillator-based Ising machines as physical learning substrates for EP~\cite{rageau2025training,11345901,ahmadi2026phase,gemo2026leveraging} with relatively sparse focus on how the underlying oscillator dynamics govern the propagation of learning signals through the network---the primary focus of this work.

To examine how the form of the oscillator interaction affects the transmission of the output perturbation during training, we consider a family of systems that encode the same Ising Hamiltonian while differing in their underlying coupling interaction. OIMs conventionally rely on phase-difference coupling, 

\begin{equation}
    \dot{\theta}_i = -K \sum_{j \neq i} J_{ij}\,\sin(\theta_i - \theta_j) - K_s \sin(2\theta_i)
\end{equation}
where $K$ is the hidden–output coupling strength,  $\theta_i$ and $\theta_j$ denote the phases of oscillator $i$ and one of its neighbors, respectively, $J_{ij}$ is the Ising couplings, and $K_s$ is the strength of SHI. In contrast, the recently proposed dynamical Ising machine (DIM) uses phase-sum coupling wherein 
\begin{equation}
\dot{\theta}_i = -K \sum_{j \neq i} J_{ij}\,\sin(\theta_i + \theta_j) - K_s \sin(2\theta_i)
\end{equation}
As shown by Ekanayake \emph{et al.}~\cite{ekanayake2025different}, under external second-harmonic injection, both coupling forms can realize the same Ising Hamiltonian. On the binary manifold, the two interactions become identical, since $\cos(\theta_i-\theta_j)=\cos(\theta_i+\theta_j)=s_i s_j$, where $s_i,s_j\in\{+1,-1\}$ correspond to oscillator phases locked at $0$ or $\pi$, respectively. This equivalence is preserved for a continuous interpolation between the two interaction forms,
\begin{align}
\dot{\theta}_i =
&-K\sum_{j} J_{ij}\left[(1-\alpha)\sin(\theta_i-\theta_j)
+\alpha\sin(\theta_i+\theta_j)\right] \notag \\
&-K_s\sin(2\theta_i),
\end{align}
with $\alpha\in[0,1]$. This interpolation retains the same binary Ising Hamiltonian~\cite{ekanayake2026mind}. The interpolation therefore provides a controlled family of oscillator systems with identical binary energy representations but different continuous-phase interactions. This allows us to isolate how the form of the underlying oscillator interaction affects the transmission of an output perturbation during equilibrium propagation.

We consider a bipartite network [Fig.~\ref{fig:architecture}] in which the hidden--output interaction is continuously tuned by the parameter $\alpha$, with $\alpha=0$ corresponding to pure phase-difference coupling and $\alpha=1$ to pure phase-sum coupling. We show that, for the oscillator model considered here, systems at $\alpha$ and $1-\alpha$ are related by a reflection of the output phases. Under the reflection symmetry of the output energy and loss, this mapping extends to the full dynamical trajectories, free and nudged equilibria, Hessian spectra, and EP parameter updates. The symmetry-fixed point occurs at $\alpha=1/2$, where the mixed hidden--output curvature vanishes on the binary manifold. As a result, an output perturbation can remain finite while its transmission to the hidden layer is strongly suppressed. Consistent with this picture, numerical training exhibits a broad low-accuracy region surrounding the midpoint, with performance approaching chance level near $\alpha=1/2$.

We further show that the interpolation parameter $\alpha$ and the strength of SHI control distinct aspects of this process. Increasing the SHI strength ($K_s$) drives the oscillator phases closer to the binary values $0$ and $\pi$~\cite{farasat2026spin}, thereby reducing the phase deviation required to generate a finite output nudging force. In contrast, $\alpha$ primarily controls the cross-layer interaction through which that output perturbation is transmitted to the hidden layer. Thus, SHI predominantly regulates the strength of the applied learning perturbation, whereas $\alpha$ regulates its propagation across layers.

\section{Oscillator Network Model and Equilibrium Propagation}
\label{sec:model}

\subsection{Network energy and phase dynamics}
In this section, we formulate the total network energy and identify the contributions from the hidden layer, output layer, and hidden--output interaction. We consider an oscillator network composed of a hidden layer with oscillator phases,
$\bm{u}=(u_1,\ldots,u_{n_h})$, and an output layer with phases $\bm{v}=(v_1,\ldots,v_{n_y})$. The real-valued input $\bm{x}$ is clamped rather than represented by an additional dynamical layer and provides synaptic drive to the hidden layer. With
\begin{equation}
  \bm{z}=\begin{bmatrix}\bm{u}\\ \bm{v}\end{bmatrix},
\end{equation}
the total network energy can be expressed as:
\begin{equation}
  E_\alpha(\bm{z};\bm{x},\vartheta)
  =E_h(\bm{u};\bm{x},\vartheta)
  +E_y(\bm{v};\vartheta)
  +E_{\mathrm{coup},\alpha}(\bm{u},\bm{v};\vartheta),
  \label{eq:physical_energy}
\end{equation}
where $\vartheta$ denotes the trainable couplings and biases. Here, $E_{\mathrm{coup},\alpha}(\bm{u},\bm{v};\vartheta)$ can be expressed as:
\begin{align}
  E_{\mathrm{coup},\alpha}(\bm{u},\bm{v};\vartheta)
  =-K\sum_{i=1}^{n_h}\sum_{a=1}^{n_y}W^{(hy)}_{ia}
  \bigl[&(1-\alpha)\cos(u_i-v_a)
  \nonumber\\
  &+\alpha\cos(u_i+v_a)\bigr].
  \label{eq:hybrid_coupling}
\end{align}
Where $W^{(hy)}_{ia}$ is the coupling between hidden oscillator $i$ and output oscillator $a$. The limits $\alpha=0$ and $\alpha=1$ correspond to phase-difference and phase-sum interactions, respectively. Furthermore, the local energies of each layer can be defined as,
\begin{align}
  E_h(\bm{u};\bm{x},\vartheta)
  &=-\sum_i h_i^{(h)}(\bm{x})\cos u_i
    -\frac{K_{s,h}}{2}\sum_i\cos(2u_i),
    \label{eq:local_energies1}
  \\
  E_y(\bm{v};\vartheta)
  &=-\sum_a b_a^{(y)}\cos v_a
    -\frac{K_{s,y}}{2}\sum_a\cos(2v_a),
  \label{eq:local_energies}
\end{align}
where
\begin{equation}
  h_i^{(h)}(\bm{x})=b_i^{(h)}+\sum_j W^{(xh)}_{ji}x_j.
  \label{eq:input_bias}
\end{equation}
where $W^{(xh)}_{ji}$ is the weight from input $j$ to hidden oscillator $i$; $b^{(h)}_i$ and $b^{(y)}_a$ are the hidden and output biases, and $K_{s,h}$ and $K_{s,y}$ denote the strengths of the SHI for the hidden-layer and output-layer oscillators, respectively.

At a binary configuration, $u_i,v_a\in\{0,\pi\}$, define
$s_i=\cos u_i$ and $s_a=\cos v_a$. Since
$\cos(u_i-v_a)=\cos(u_i+v_a)=s_is_a$, the interlayer coupling energy  reduces to
\begin{equation}
  E_{\mathrm{coup},\alpha}^{0/\pi}
  =-K\sum_{i,a}W^{(hy)}_{ia}s_is_a,
  \label{eq:binary_forward_energy}
\end{equation}
which is independent of $\alpha$. Thus every member of the interpolation
implements the same pairwise Ising energy on the binary manifold. 

EP also relies on a small output nudge to displace the network away from its free equilibrium, and the resulting phase response determines the learning signal. We therefore next characterize the force generated by this output perturbation and how it propagates through the network.

The output is encoded through $\cos v_a$, with $\hat y_a$ denoting the corresponding target value. We define the nudging loss as
\begin{equation}
\ell(\bm{v},\hat{\bm{y}})
=\sum_a\left(\cos v_a-\hat y_a\right)^2.
\label{eq:loss}
\end{equation}
Thus, the total network energy function during the nudge phase is
\begin{equation}
  F_\alpha(\bm{z};\beta,\vartheta)
  =E_\alpha(\bm{z};\bm{x},\vartheta)
  +\beta\,\ell(\bm{v},\hat{\bm{y}}),
  \label{eq:full_energy}
\end{equation}
where $\beta$ is the output-nudging strength. In the gradient descent framework,  the resulting dynamics can be defined as,
\begin{equation}
  \dot{\bm{z}}=-\nabla_{\bm{z}}F_\alpha.
  \label{eq:dynamics}
\end{equation}
During the free phase, $\beta=0$, and the dynamics reduce to gradient descent on the network energy,
\begin{equation}
\dot{\mathbf z}=-\nabla_{\mathbf z}E_\alpha .
\end{equation}
Consequently, the energy evolves over time according to
\begin{equation}
\dot{E}_\alpha 
=
\nabla_{\mathbf z}E_\alpha\cdot\dot{\mathbf z}
=
-\left\|\nabla_{\mathbf z}E_\alpha\right\|^2
\leq 0,
\end{equation}
and therefore decreases monotonically along the trajectory. Since $E_\alpha$ is non-increasing along the gradient-flow trajectory, the network relaxes toward a stationary state satisfying
\(
\nabla_{\bm z}E_\alpha\!\left(\bm z_\alpha^{*(0)}\right)=0,
\)
which we denote as the free equilibrium $\bm z_\alpha^{*(0)}$. In the nudged phase, $\beta>0$, the loss term introduces an additional force on the output phases that drives them toward the target phase, causing the network to relax to a nearby perturbed equilibrium $\mathbf z_\alpha^{*(\beta)}$. Equilibrium propagation extracts the learning signal from the difference between these two equilibria, with the displacement $\mathbf z_\alpha^{*(\beta)} -\,\mathbf z_\alpha^{*(0)}$ characterizing the response of the network to the output perturbation.

\subsection{Linear response to output nudging}
\noindent We now derive how the equilibrium shifts when a weak nudge is applied at the outputs, and show that this shift determines the EP parameter update during the training.
For fixed input and parameters, the free equilibrium $\boldsymbol{z}_\alpha^{*(0)}$ satisfies
\begin{equation}
\nabla_{\boldsymbol{z}} E_\alpha\big(\boldsymbol{z}_\alpha^{*(0)}\big) = 0.
\label{eq:free-eq}
\end{equation}
The nudged objective is
\begin{equation}
F(\boldsymbol{z},\beta;\vartheta) \;=\; E_\alpha(\boldsymbol{z};\vartheta) \;+\; \beta\,\ell\big(\boldsymbol{z},\hat y\big).
\label{eq:nudged-objective}
\end{equation}
The nudged equilibrium $\boldsymbol{z}_\alpha^{*(\beta)}$ is a stationary point of $F$, i.e.\ it satisfies
\begin{equation}
\nabla_{\boldsymbol{z}} E_\alpha\big(\boldsymbol{z}_\alpha^{*(\beta)};\vartheta\big)
\;+\; \beta\,\nabla_{\boldsymbol{z}} \ell\big(\boldsymbol{z}_{\alpha}^{*(\beta)},\hat y\big) \;=\; 0.
\label{eq:nudged-eq}
\end{equation}
For sufficiently small $\beta$, the implicit function theorem guarantees a unique differentiable function $f(\beta) = \boldsymbol{z}_\alpha^{*(\beta)}$ and its Taylor series expansion around $\beta=0$ gives
\begin{equation}
\boldsymbol{z}_\alpha^{*(\beta)} \;=\; \boldsymbol{z}_\alpha^{*(0)} \;+\; \beta\,\delta \boldsymbol{z}_\alpha \;+\; \mathcal{O}(\beta^2).
\label{eq:ift-expansion}
\end{equation}
Substituting the expansion in Eq.~\eqref{eq:ift-expansion} into the nudged equilibrium condition of Eq.~\eqref{eq:nudged-eq} yields,
\begin{align}
\nabla_{\boldsymbol{z}} E_\alpha\big(\boldsymbol{z}_\alpha^{*(0)};\vartheta\big)
\;+\; \beta\,\nabla^2_{\boldsymbol{z}\boldsymbol{z}} E_\alpha\big(\boldsymbol{z}_\alpha^{*(0)};\vartheta\big)\,\delta \boldsymbol{z}_\alpha
\;
\nonumber\\
+\; \beta\,\nabla_{\boldsymbol{z}} \ell\big(\boldsymbol{z}_{\alpha}^{*(0)},\hat y\big)
\;+\; \mathcal{O}(\beta^2) \;=\; 0.
\label{eq:linearized}
\end{align}
The zeroth-order term $\nabla_{\boldsymbol{z}} E_\alpha(\boldsymbol{z}_\alpha^{*(0)};\vartheta)$ vanishes by Eq.~\eqref{eq:free-eq}. Dividing the remaining $\mathcal{O}(\beta)$ terms by $\beta$ and using the Hessian definition,
\begin{equation}
  H_\alpha
  =\nabla_{\bm{z}\bm{z}}^2E_\alpha
  (\bm{z}_\alpha^{*(0)};\vartheta)
  \label{eq:hessian_definition}
\end{equation}
we obtain
\begin{equation}
H_\alpha\,\delta \boldsymbol{z}_\alpha \;+\; \boldsymbol{q} \;\approx\; 0, \qquad\,\nabla_{\bm{z}}\ell\bigl(\bm{z}^{*(0)}_\alpha\bigr) = \,\bm{q},
\label{eq:linear-system}
\end{equation}
and hence,
\begin{equation}
{\;\delta \boldsymbol{z}_\alpha \;\approx\; -H_\alpha^{-1} \boldsymbol{q}.\;}
\label{eq:delta-z-solution}
\end{equation}
Here, $H_\alpha^{-1}$ determines how the applied nudging force $\boldsymbol q$ produces a displacement of the oscillator phases, $\delta\boldsymbol z_\alpha$. Equation~(\ref{eq:delta-z-solution}) therefore separates the linear response into two factors: the force generated by the output nudge and the response of the network state to that force. The latter is characterized by $H_\alpha^{-1}$, which we refer to as the local susceptibility of the network. Since $H_\alpha$ is the Hessian of the physical energy evaluated at the free equilibrium, this susceptibility is determined by the local curvature of the energy landscape.

The training objective is the minimization of the loss evaluated at the free equilibrium,
\begin{equation}
  L(\vartheta) = \ell\bigl(\bm{v}^{*(0)}_\alpha, \hat{\bm{y}}\bigr).
  \label{eq:objective}
\end{equation}
As Eq.~\eqref{eq:objective} shows, the loss has no direct dependence on $\vartheta$. As a result, the loss affects $\vartheta$ only through the equilibrium shift, and the gradient in $\vartheta$ is only dependent on that shift. The one-sided EP
gradient estimator (see Appendix~\ref{app:ep_gradient}) is 
\begin{equation}
  \widehat{\nabla_\vartheta L}
  = \frac{1}{\beta}\Bigl[\partial_\vartheta E_\alpha\bigl(\bm{z}^{*(\beta)}_\alpha\bigr)
    - \partial_\vartheta E_\alpha\bigl(\bm{z}^{*(0)}_\alpha\bigr)\Bigr].
  \label{eq:ep_estimator}
\end{equation}
Using Eq.~\eqref{eq:ift-expansion},
\begin{equation}
  \widehat{\nabla_\vartheta L}
  = \bigl(\nabla^2_{\bm{z}\vartheta} E_\alpha\bigr)^{\mathsf{T}} \delta\bm{z}_\alpha
    + \mathcal{O}(\beta).
  \label{eq:ep_gradient_linear}
\end{equation}
As a result, the same physical response that connects the free and nudged equilibria also sets the leading EP parameter gradient.

\section{Output-to-hidden susceptibility and learning-signal propagation}
\label{sec:response}

\subsection{Decomposition of the output-to-hidden response}
\noindent We now decompose the linear response $\delta\bm{z}_\alpha$ into its hidden and output components to identify the factors that control how the nudge force is transmitted to the hidden layer.
Since the loss acts only on the output variables, the loss gradient has the block form
\begin{equation}
  \bm{q}=
  \begin{bmatrix}
    \textbf{0}\\ \bm{q}_y
  \end{bmatrix}.
  \label{eq:loss_force_block}
\end{equation}
Partitioning the free-state Hessian as
\begin{equation}
  H_\alpha=
  \begin{bmatrix}
    H_{hh} & H_{hy}\\
    H_{yh} & H_{yy}
  \end{bmatrix},
  \label{eq:hessian_blocks}
\end{equation}
where $H_{hh}=\nabla^2_{uu}E_\alpha$ and $H_{yy}=\nabla^2_{vv}E_\alpha$ are the hidden- and output-layer curvature blocks, and $H_{hy}=\nabla^2_{uv}E_\alpha=H_{yh}^{\mathsf{T}}$ is the mixed hidden–output curvature. The linear-response equation can then be expressed as,
\begin{equation}
  \begin{bmatrix}
    H_{hh} & H_{hy}\\
    H_{yh} & H_{yy}
  \end{bmatrix}
  \begin{bmatrix}
    \delta\bm{u}\\ \delta\bm{v}
  \end{bmatrix}
  =-
  \begin{bmatrix}
    \textbf{0}\\ \bm{q}_y
  \end{bmatrix}.
  \label{eq:block_response}
\end{equation}
Eliminating the hidden variables gives the output Schur complement
\begin{equation}
  S_y=H_{yy}-H_{yh}H_{hh}^{-1}H_{hy},
  \label{eq:schur}
\end{equation}
and the responses
\begin{align}
  \delta\bm{v}&=-S_y^{-1}\bm{q}_y,
  \label{eq:output_response}\\
  \delta\bm{u}&=H_{hh}^{-1}H_{hy}S_y^{-1}\bm{q}_y
  \equiv T_\alpha\bm{q}_y.
  \label{eq:hidden_response}
\end{align}
The operator
\begin{equation}
  T_\alpha=H_{hh}^{-1}H_{hy}S_y^{-1}
  \label{eq:propagation_operator}
\end{equation}
is the physical output layer-to-hidden layer susceptibility. It separates the transmission
of the nudge force into three components: the output layer response $S_y^{-1}$, the mixed
hidden--output curvature $H_{hy}$, and the hidden layer response $H_{hh}^{-1}$. If the hidden phases were held fixed, the output response to the applied nudge would be determined by $H_{yy}^{-1}$. However, since the hidden phases are also free to evolve, their response to the output perturbation modifies the effective curvature of the output layer. This effect is captured by the Schur complement $S_y=H_{yy}-H_{yh}H_{hh}^{-1}H_{hy}$. Consequently, the output displacement is governed by $S_y^{-1}$ rather than $H_{yy}^{-1}$, accounting for the coupled response of the hidden and output oscillators.

Equation~\eqref{eq:hidden_response} shows that the hidden-layer response depends on both the applied nudging force $\bm q_y$ and the output-to-hidden susceptibility $T_\alpha$. The output displacement $\delta\bm v=-S_y^{-1}\bm q_y$ generates a perturbation in the hidden layer through the mixed curvature $H_{hy}$, while $H_{hh}^{-1}$ determines the resulting hidden-phase displacement. Consequently, an appreciable output displacement does not necessarily produce an appreciable hidden-layer response. In particular, when $H_{hy}$ is small and the remaining susceptibility factors are bounded, the transmission of the output perturbation to the hidden layer is suppressed.

In the oscillator network considered here, the interpolation parameter $\alpha$ directly controls the mixed curvature $H_{hy}$. At a fixed binary phase configuration, the diagonal Hessian blocks are independent of $\alpha$, whereas $H_{hy}$ is proportional to $2\alpha-1$. Thus, at $\alpha=1/2$, the mixed curvature vanishes on the binary manifold, eliminating the linear coupling between output and hidden phase perturbations. The behavior near this point is examined in Sec.~\ref{sec:binary}.

\subsection{Influence of susceptibility on EP learning}

We now examine how the output-to-hidden susceptibility $T_\alpha$ influences equilibrium propagation by determining the learning signals received by different groups of trainable parameters. As established in Eq.~\eqref{eq:ep_gradient_linear}, the leading-order EP gradient depends on the displacement of the oscillator phases between the free and nudged equilibria. Since different parameters contribute to different terms in the network energy, their updates depend on the responses of different oscillator layers.

Consider first the parameters $\vartheta_h$ that enter exclusively through the hidden-layer energy $E_h$, including the hidden biases and input-to-hidden coupling weights defined in Eqs.~\eqref{eq:local_energies1} and~\eqref{eq:input_bias}. For these parameters, Eq.~\eqref{eq:ep_gradient_linear} reduces to
\begin{equation}
  \widehat{\nabla_{\vartheta_h} L}
  =
  \bigl(\nabla^2_{\bm{u}\vartheta_h} E_h\bigr)^{\mathsf{T}}
  \delta\bm{u}
  + \mathcal{O}(\beta).
  \label{eq:hidden_param_gradient}
\end{equation}
Thus, the leading-order learning signal for these parameters depends entirely on the hidden-layer displacement $\delta\bm{u}$. Since $\delta\bm{u}=T_\alpha\bm{q}_y$, a small output-to-hidden susceptibility suppresses this displacement for a bounded nudging force, thereby weakening the updates to the input-to-hidden weights and hidden biases. Importantly, this suppression can occur even when the output oscillators undergo an appreciable displacement in response to the applied nudge.

The hidden-to-output coupling weights behave differently because their energy contribution depends on the phases of both layers. Consequently, their EP gradients contain contributions from both the hidden- and output-layer displacements:
\begin{align}
  \widehat{\nabla_W L}
  = {}&
  \bigl(\nabla^2_{\bm{u}W} E_{\mathrm{coup},\alpha}\bigr)^{\mathsf{T}}
  \delta\bm{u}
  \nonumber\\
  &+
  \bigl(\nabla^2_{\bm{v}W} E_{\mathrm{coup},\alpha}\bigr)^{\mathsf{T}}
  \delta\bm{v}
  + \mathcal{O}(\beta).
  \label{eq:W_gradient}
\end{align}
The first term represents the contribution from the hidden-layer response, whereas the second arises directly from the displacement of the output oscillators. Therefore, even when $\delta\bm{u}$ is strongly suppressed, the hidden-to-output weights can retain a learning signal through $\delta\bm{v}$, provided the corresponding energy derivatives remain nonzero. Similarly, parameters associated exclusively with the output-layer energy depend on the output displacement and are not directly limited by the output-to-hidden susceptibility.

These results show that suppression of the output-to-hidden susceptibility does not necessarily eliminate learning throughout the network. Instead, it selectively inhibits the updates of parameters whose learning signals depend on the hidden-layer response, while parameters coupled directly to the output phases may continue to receive learning signals.

\section{Suppression of Hidden-Layer Response Near the Binary Manifold}
\label{sec:binary}

\subsection{Mixed Hidden-Output Curvature}
We now examine the hidden--output curvature near the binary manifold and establish the conditions under which it vanishes at the symmetry point $\alpha=1/2$. Defining $\gamma=2\alpha-1$, the mixed second derivative of the network energy in Eq.~\eqref{eq:hybrid_coupling} gives
\begin{equation}
\begin{split}
[H_{hy}]_{ia}
&=\frac{\partial^2 E_\alpha}{\partial u_i\,\partial v_a}\\
&=-K W^{(hy)}_{ia}
\left[
\sin u_i\sin v_a
-\gamma\cos u_i\cos v_a
\right].
\end{split}
\label{eq:mixed_hessian_general}
\end{equation}
We consider two limiting cases: binary oscillator phases at arbitrary $\alpha$, and the symmetry point $\alpha=1/2$ at arbitrary phases.

At a binary configuration, $u_i=P_i\pi$ and $v_a=Q_a\pi$, with $P_i,Q_a\in\mathbb{Z}$, so the sine terms vanish. Defining $s_i=\cos u_i$ and $s_a=\cos v_a$, Eq.~\eqref{eq:mixed_hessian_general} reduces to
\begin{equation}
[H_{hy}]_{ia}
=
K\gamma W^{(hy)}_{ia}s_is_a.
\label{eq:mixed_hessian_binary}
\end{equation}
Thus, the mixed hidden--output curvature is proportional to $2\alpha-1$ and vanishes at $\alpha=1/2$. At this point, the Hessian becomes block diagonal on the binary manifold, eliminating the linear coupling between hidden- and output-phase perturbations.

We next consider $\alpha=1/2$ without restricting the oscillator phases to binary values. At this point, $\gamma=0$, and the coupling energy in Eq.~\eqref{eq:hybrid_coupling} factorizes as
\begin{equation}
E_{\mathrm{coup},1/2}
=
-K\sum_{i,a}W^{(hy)}_{ia}\cos u_i\cos v_a.
\label{eq:midpoint_factorization}
\end{equation}
The corresponding mixed curvature becomes
\begin{equation}
[H_{hy}]_{ia}
=
-KW^{(hy)}_{ia}\sin u_i\sin v_a.
\label{eq:midpoint_mixed}
\end{equation}
Unlike the binary case, this curvature is generally nonzero. However, as the oscillator phases approach $0$ or $\pi$, both sine factors become small, causing the mixed curvature to decrease. Consequently, the output-to-hidden response is strongly suppressed near the binary manifold, provided the remaining susceptibility factors remain bounded. Importantly, the coupling energy in Eq.~\eqref{eq:midpoint_factorization} remains finite, demonstrating that suppression of the mixed curvature does not imply elimination of the forward interaction.

\subsection{Vanishing Nudging Force at Binary Phases}

For the loss defined in Eq.~\eqref{eq:loss}, the gradient with respect to the output phases is
\begin{equation}
  q_{y,a} = \frac{\partial \ell}{\partial v_a}
  = -2\left(\cos v_a - \hat{y}_a\right)\sin v_a,
  \label{eq:loss_force}
\end{equation}
which produces a nudging force $-\beta q_{y,a}$ on output oscillator $a$. At binary phases, $v_a\in\{0,\pi\}$, the factor $\sin v_a$ vanishes, making the nudging force zero regardless of the difference between the predicted and target outputs. Consequently, $\bm q=\bm 0$, and Eq.~\eqref{eq:delta-z-solution} gives $\delta\bm z=\bm 0$, provided the free-state Hessian is non-singular. Thus, an exactly binary configuration produces no first-order response to output nudging, irrespective of $\alpha$.

\subsection{Near-binary scaling of the hidden layer response}
\label{sec:nearbinary}

We also explore the response of the hidden layer in the vicinity of the binary manifold beyond the exact binary limit. We express the phases as small deviations from $0$ or $\pi$:
\begin{equation}
  u_i = P_i\pi + \epsilon_i,
  \qquad
  v_a = Q_a\pi + \zeta_a,
  \label{eq:near_binary_coordinates}
\end{equation}
with $|\epsilon_i|, |\zeta_a| \ll 1$. Expanding
Eq.~\eqref{eq:mixed_hessian_general} to leading orders gives
\begin{widetext}
\begin{equation}
  [H_{hy}]_{ia}
  = -K W^{(hy)}_{ia} s_i s_a
  \left[
    \epsilon_i\zeta_a - \gamma
    + \frac{\gamma}{2}\left(\epsilon_i^2 + \zeta_a^2\right)
  \right]
  + \mathcal{O}\!\left(\epsilon_i^3\zeta_a,\ \epsilon_i\zeta_a^3,\
  |\gamma|\epsilon_i^2\zeta_a^2,\ \epsilon_i^3\zeta_a^3\right).
  \label{eq:near_binary_mixed}
\end{equation}
\end{widetext}
It can be observed that Eq.~\eqref{eq:near_binary_mixed}  contains a leading contribution proportional to $\gamma=2\alpha-1$ and a contribution proportional to the product of the phase deviations, $\epsilon_i\zeta_a$. Since $|\epsilon_i|,|\zeta_a|\ll1$, the latter is of second order in the phase deviations. Thus, when $|\gamma|\gg|\epsilon_i\zeta_a|$, the $\gamma$-dependent term dominates, giving $H_{hy}=\mathcal{O}(|\gamma|)$. At the symmetry point $\alpha=1/2$, however, $\gamma=0$, and the mixed curvature is determined entirely by the smaller contribution proportional to $\epsilon_i\zeta_a$.

The nudging force is also suppressed near the binary manifold. From Eq.~\eqref{eq:loss_force}, $\sin v_a=\mathcal{O}(\zeta_a)$, so the output loss gradient satisfies $\bm q_y=\mathcal{O}(\epsilon_y)$, where $\epsilon_y=\max_a|\zeta_a|$. Combining these results with Eq.~\eqref{eq:hidden_response}, and assuming that the inverse curvature matrices $H_{hh}^{-1}$ and $S_y^{-1}$ remain bounded in the regime considered, we obtain
\begin{equation}
    \|\delta\bm{u}\|
    =
    \mathcal{O}\!\left[
    \left(|\gamma|+\epsilon_h\epsilon_y\right)\epsilon_y
    \right],
    \label{eq:du_scaling}
\end{equation}
where $\epsilon_h=\max_i|\epsilon_i|$ denotes the maximum hidden-phase deviation from the binary manifold.

For comparable phase deviations, $\epsilon_h\sim\epsilon_y\sim\epsilon$, Eq.~\eqref{eq:du_scaling} gives a hidden-layer response of $\mathcal{O}(|2\alpha-1|\epsilon)$ when $|2\alpha-1|\gg\epsilon^2$. At the midpoint, however, the leading contribution proportional to $2\alpha-1$ vanishes, leaving a response of only $\mathcal{O}(\epsilon^3)$. Thus, the symmetry point exhibits a stronger suppression of the hidden-layer response to output nudging as the oscillator phases approach the binary manifold.

The suppression is not restricted to $\alpha=1/2$. Equation~\eqref{eq:du_scaling} shows that when $|\gamma|\lesssim\epsilon_h\epsilon_y$, the contribution from the phase deviations becomes comparable to or larger than the $\gamma$-dependent contribution, and the hidden response remains bounded by $\mathcal{O}(\epsilon_h\epsilon_y^2)$. This defines a region around the midpoint in which the transmission of output perturbations to the hidden layer is strongly suppressed. Since the leading-order EP gradients of the hidden-layer parameters $W^{(xh)}$ and $b^{(h)}$ depend on $\delta\bm{u}$ [Eq.~\eqref{eq:hidden_param_gradient}], their learning signals are correspondingly weakened. The extent of this suppression depends on both the coupling parameter $\alpha$ and the proximity of the oscillator phases to the binary manifold.

\section{Exact reflection symmetry of the oscillator dynamics}
\label{sec:conjugacy}
We show that reflecting the output phases, $\bm{v}\rightarrow-\bm{v}$, maps the oscillator system at $\alpha$ onto the system at $1-\alpha$. For the model considered here, this mapping establishes an exact equivalence of the nonlinear dynamics, with $\alpha=1/2$ as the symmetry-fixed point. This symmetry has two important consequences. First, at $\alpha=1/2$, the mixed hidden--output curvature vanishes at binary phase configurations, consistent with Eq.~\eqref{eq:mixed_hessian_binary}. This cancellation follows from the reflection symmetry of the energy. Second, the systems at $\alpha$ and $1-\alpha$ have equivalent dynamical properties, including corresponding trajectories, equilibria, and Hessian spectra. Consequently, physical observables that are invariant under output-phase reflection exhibit identical behavior at these two values of $\alpha$, provided the initial conditions are related by the same reflection.

To establish the reflection symmetry of the oscillator dynamics, we consider the transformation $\bm{v}\rightarrow-\bm{v}$, which reverses the output phases while leaving the hidden phases unchanged. For the full state vector $\bm{z}=(\bm{u},\bm{v})^{\mathsf T}$, this transformation is represented by
\begin{equation}
  R=
  \begin{bmatrix}
    I_h & 0\\
    0 & -I_y
  \end{bmatrix},
  \label{eq:reflection_matrix}
\end{equation}
where $I_h$ and $I_y$ are identity matrices corresponding to the hidden and output layers, respectively.

Under this transformation, the phase-difference and phase-sum interactions are exchanged, since $\cos(u_i-v_a)$ becomes $\cos(u_i+v_a)$ and vice versa. Consequently, the coupling energy in Eq.~\eqref{eq:hybrid_coupling} satisfies
\begin{equation}
  E_{\mathrm{coup},\alpha}(\bm{u},-\bm{v})
  =
  E_{\mathrm{coup},1-\alpha}(\bm{u},\bm{v}).
  \label{eq:coupling_reflection}
\end{equation}
Thus, reflecting the output phases is equivalent to replacing $\alpha$ with $1-\alpha$, or $\gamma$ with $-\gamma$.

The remaining terms in the network energy are invariant under this reflection. In particular, the output-layer energy in Eq.~\eqref{eq:local_energies} and the loss in Eq.~\eqref{eq:loss} depend only on $\cos v_a$ and $\cos 2v_a$, which are unchanged under $v_a\rightarrow-v_a$. The hidden-layer energy is also unaffected because it does not depend on the output phases. The complete nudged energy therefore obeys
\begin{equation}
  F_\alpha(R\bm{z};\beta,\vartheta)
  =
  F_{1-\alpha}(\bm{z};\beta,\vartheta).
  \label{eq:energy_conjugacy}
\end{equation}
We note that this identity is not restricted to the binary manifold and  holds for arbitrary oscillator phases as well as finite nudging strength $\beta$.

Since $R=R^{\mathsf T}=R^{-1}$, Eq.~\eqref{eq:energy_conjugacy} can equivalently be written as
\begin{equation}
  F_{1-\alpha}(R\bm{z};\beta,\vartheta)
  =
  F_\alpha(\bm{z};\beta,\vartheta).
\end{equation}
Differentiating this identity with respect to $\bm{z}$ and applying the chain rule yields
\begin{align}
  \nabla_{\bm{z}}F_{1-\alpha}(R\bm{z})
  &=
  R\nabla_{\bm{z}}F_\alpha(\bm{z}),
  \label{eq:grad_conjugacy}\\
  \nabla_{\bm{z}}^2F_{1-\alpha}(R\bm{z})
  &=
  R\nabla_{\bm{z}}^2F_\alpha(\bm{z})R.
  \label{eq:hess_conjugacy}
\end{align}
Here, the dependence on $\beta$ and $\vartheta$ has been omitted for brevity.

Equation~\eqref{eq:grad_conjugacy} establishes the equivalence of the full gradient-flow dynamics. If $\bm{z}_\alpha(t)$ is a trajectory at $\alpha$, then
\begin{equation}
\begin{split}
  \frac{d}{dt}\left(R\bm{z}_\alpha\right)
  &=
  -R\nabla_{\bm{z}}F_\alpha(\bm{z}_\alpha)\\
  &=
  -\nabla_{\bm{z}}F_{1-\alpha}
  (R\bm{z}_\alpha).
\end{split}
\end{equation}
Thus, $R\bm{z}_\alpha(t)$ is a trajectory of the system at $1-\alpha$, provided the initial conditions are related by the same reflection. The free and nudged equilibria reached from these corresponding initial conditions consequently satisfy
\begin{equation}
  \bm{z}_{1-\alpha}^{*(\beta)}
  =
  R\bm{z}_{\alpha}^{*(\beta)}.
  \label{eq:equilibria_conjugacy}
\end{equation}

The Hessian identity in Eq.~\eqref{eq:hess_conjugacy} further implies that the Hessians evaluated at corresponding equilibria are related by
\begin{equation}
  H_{1-\alpha}
  =
  RH_\alpha R^{-1}.
\end{equation}
Since this is an orthogonal similarity transformation, the Hessians have identical eigenvalue spectra and spectral condition numbers. The corresponding equilibria therefore possess identical local stability properties.

Finally, this equivalence extends to the EP parameter updates. Since the reflection identity holds for the same trainable parameters $\vartheta$, differentiation of the physical energy with respect to these parameters gives
\begin{equation}
  \partial_\vartheta E_{1-\alpha}(R\bm{z})
  =
  \partial_\vartheta E_\alpha(\bm{z}).
  \label{eq:update_conjugacy}
\end{equation}
Together with Eq.~\eqref{eq:equilibria_conjugacy}, this shows that the differences in energy derivatives between corresponding free and nudged equilibria are identical. Consequently, the paired systems generate the same EP parameter updates, including those for the hidden--output coupling weights and the hidden- and output-layer biases.

The reflection therefore establishes an exact equivalence between the systems at $\alpha$ and $1-\alpha$, encompassing their nonlinear trajectories, equilibria, Hessian spectra, and EP learning updates.

\section{Simulations}
\label{sec:numerics}
We now investigate the theoretical results developed above, focusing on the exact reflection symmetry between $\alpha$ and $1-\alpha$, the distinct roles of the nudging force and cross-layer curvature, and the suppression of the hidden-layer response near the binary manifold.

\subsection{Numerical setup}
\label{sec:setup}

We consider a bipartite oscillator network with architecture $784\rightarrow 200\rightarrow 10$, consisting of 784 input variables, $n_h=200$ hidden oscillators, and $n_y=10$ output oscillators. The network is trained and evaluated on the MNIST handwritten-digit dataset~\cite{lecun1998gradient}, containing $60,000$ training images and $10,000$ test images, each with $28\times28$ grayscale pixels. The input is held fixed during oscillator relaxation and enters through the hidden-layer fields defined in Eq.~\eqref{eq:input_bias}. The target labels are encoded as bipolar one-hot vectors, with $+1$ assigned to the correct class and $-1$ to all other classes. The predicted class corresponds to the output oscillator with the largest value of $\cos v_a$.

The input-to-hidden and hidden-to-output weights are initialized using the variance-scaling scheme of Ref.~\cite{he2015delving}:
\begin{equation}
W'^{(xh)}_{ji} \sim \mathcal{N}\left(0,\frac{2}{n_x}\right),
\qquad
W'^{(hy)}_{ia} \sim \mathcal{N}\left(0,\frac{2}{n_h}\right),
\label{weight}
\end{equation}
where $W'$ is before rescaling,  and $n_x=784$. All bias parameters are initialized to zero.

The input pixel intensities are rescaled from $x_j\in[0,1]$ to $x'_j=2x_j-1\in[-1,1]$ to obtain a bipolar representation. This transformation changes the magnitude of the input-dependent hidden fields because background pixels, originally close to zero, contribute values close to $-1$ after rescaling. For the initialization in Eq.~\eqref{weight}, the variance of the input-induced hidden field, evaluated over the random weights for a fixed input $\bm{x}$, is
\begin{equation}
\operatorname{Var}\left(h_i^{(h)}\right)
=
\sum_j x_j^2
\operatorname{Var}\left(W'^{(xh)}_{ji}\right)
=
\frac{2}{n_x}\|\bm{x}\|^2.
\end{equation}
Thus, the typical magnitude of the hidden field depends on the norm of the input vector, which changes under the bipolar transformation.

To compensate for this change, the input-to-hidden weights are rescaled once at initialization according to
\begin{equation}
W^{(xh)}=cW'^{(xh)},
\qquad
c=
\sqrt{
\frac{\mathbb{E}\left[\|\bm{x}\|^2\right]}
{\mathbb{E}\left[\|\bm{x}'\|^2\right]}
},
\label{eq:input_normalization}
\end{equation}
where the expectations are evaluated over the training dataset. This normalization preserves the dataset-averaged variance of the initial input-induced hidden fields relative to the original $[0,1]$ encoding.

Maintaining a comparable input-field magnitude is important because the phase gradient of the hidden-layer energy contains contributions from both the input-dependent field and second-harmonic injection (SHI):
\begin{equation}
\frac{\partial E_h}{\partial u_i}
=
h_i^{(h)}\sin u_i
+
K_{s,h}\sin(2u_i).
\end{equation}
The first term represents the input-dependent contribution, whereas the second favors phase configurations near $0$ and $\pi$. Rescaling the input-to-hidden weights therefore helps maintain a comparable initial balance between these contributions when the input encoding is changed.

For all simulations, the hidden--output coupling strength is fixed at $K=1$, while the SHI strengths are varied as specified in the individual experiments. The oscillator dynamics are integrated using the explicit Euler method with a time step of $\Delta t=0.2$. Each training example undergoes $30,000$ integration steps during the free phase, followed by $10,000$ steps during the nudged phase, initialized from the free equilibrium. The nudging strength is fixed at $\beta=0.1$.

Training is performed using minibatches of 128 samples, with separate learning rates for different parameter groups. The input-to-hidden weights use $\eta_{W'^{(xh)}}=0.001$, while the hidden-to-output weights and the hidden- and output-layer biases use $\eta_{W'^{(hy)}}=\eta_{b^{(h)}}=\eta_{b^{(y)}}=0.002$.

\begin{table}[t]

\small
\begin{tabular}{lcc}
\toprule
Quantity & $\alpha = 0$ & $\alpha = 1/2$ \\
\midrule
$\max_t \|H_{hy}\|_2$          & $\approx 2$            & $0$ \\
$\max_t \|\bm{q}_y\|_2$        & $3.219$                & $1\times10^{-1}$ \\
$\max_t \|\delta\bm{u}\|$ & $8.344\times10^{-1}$   & $<{10}^{-10}$ \\
\bottomrule

\end{tabular}
\caption{\justifying Comparison of the hidden layer response to the output perturbation together with the quantities that govern it (Eq.~\eqref{eq:hidden_response}), for
$\alpha = 0$ and $\alpha = 0.5$ at $K_s = 0.1$. Maximum values over ten training epochs with identical
untrained initial weights are reported. At the midpoint, the output
is still perturbed, but this perturbation is not transmitted toward the hidden layer.}
\label{tab:peaks}
\end{table}

\subsection{Suppression of hidden–output cross-layer curvature at $\alpha=1/2$}
We numerically examine the suppression of the mixed hidden--output curvature $H_{hy}$ at $\alpha=1/2$. Table~\ref{tab:peaks} reports the maximum values of $\|H_{hy}\|$, the output nudging term $\|\bm q_y\|$, and the hidden-layer response $\|\delta\bm u\|$ over ten training epochs for $\alpha=0$ and $\alpha=1/2$.

At $\alpha=0$, the mixed curvature reaches $\|H_{hy}\|\approx 2$, and the hidden-layer response is appreciable, reaching $\|\delta\bm u\|\approx 0.83$. In contrast, at $\alpha=1/2$, the nudging term remains nonzero, with $\|\bm q_y\|\sim 10^{-1}$, while both $\|H_{hy}\|$ and $\|\delta\bm u\|$ become numerically negligible. Thus, even though the output layer experiences a finite nudging force, the resulting perturbation is not appreciably transmitted to the hidden layer. These results are consistent with the theoretical prediction that suppression of the mixed curvature can inhibit output-to-hidden perturbation transmission independently of the magnitude of the applied nudging force.

\subsection {Controlling the nudging force and the cross-layer curvature}

\begin{figure*}
    \centering

              \centering
              \includegraphics[width=1.04\linewidth]{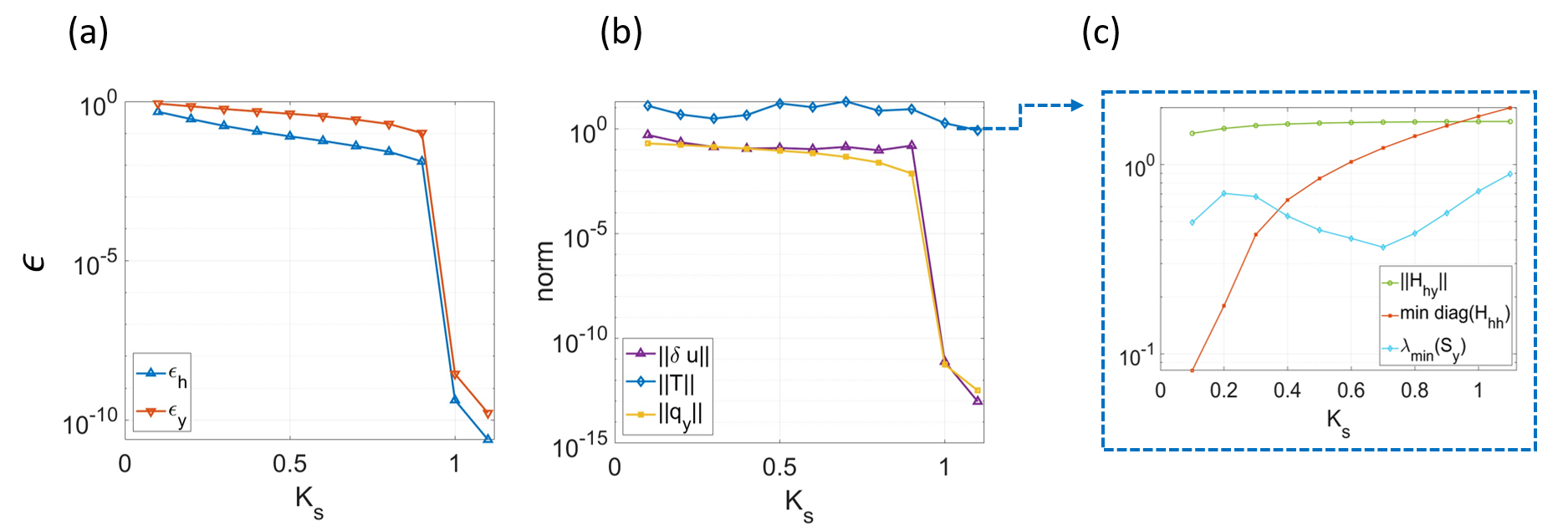}

        \caption{\justifying \small Effect of SHI strength $K_s$ on the output-to-hidden response at $\alpha=0$.
(a) Phase deviations from the binary manifold in the hidden ($\epsilon_h$) and output ($\epsilon_y$) layers as functions of $K_s$. Both decrease with increasing $K_s$, indicating stronger phase confinement near $0$ and $\pi$.
(b) Norms of the output nudging term $\|\bm{q}_y\|$, susceptibility operator $\|T_\alpha\|$, and hidden-layer response $\|\delta\bm{u}\|$ versus $K_s$. The nudging term and hidden response decrease substantially, whereas the susceptibility varies comparatively weakly.
(c) Components governing the susceptibility $T_\alpha=H_{hh}^{-1}H_{hy}S_y^{-1}$, characterized by the mixed curvature $\|H_{hy}\|$, minimum hidden-layer curvature $\min\operatorname{diag}(H_{hh})$, and minimum eigenvalue $\lambda_{\min}(S_y)$ of the output Schur complement. The mixed curvature remains nearly constant, while changes in the hidden- and output-layer curvature contribute to the comparatively small variation in $\|T_\alpha\|$.}

\label{fig:sussweep}
\end{figure*}

\begin{figure*}
    \centering
    \includegraphics[width=1.00\linewidth]{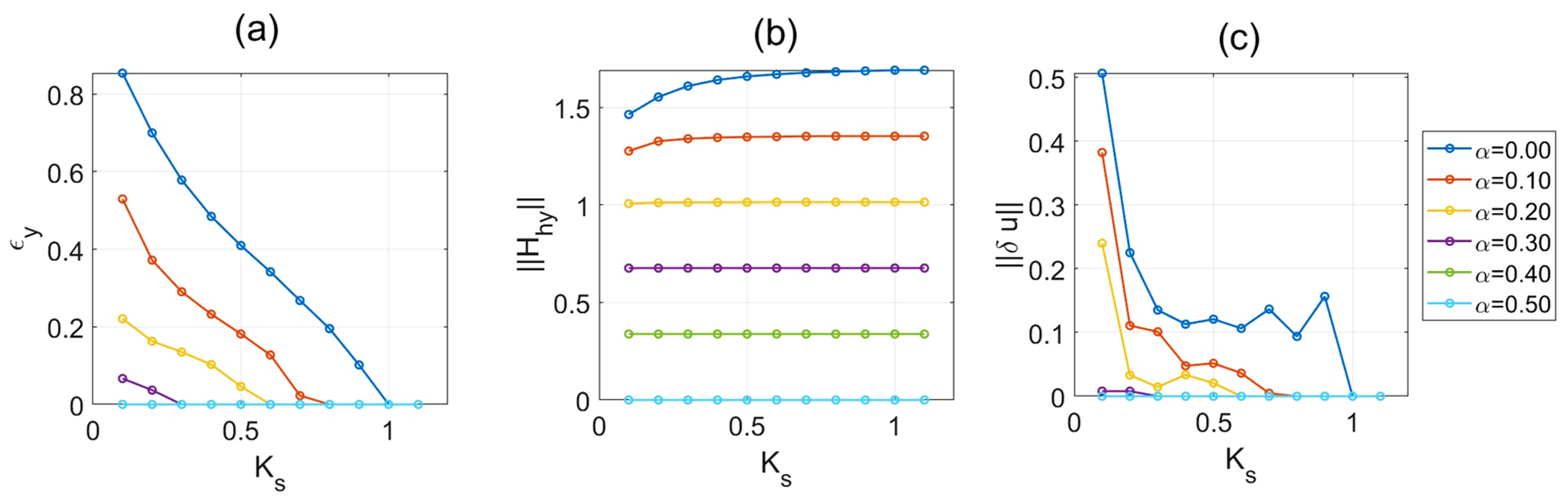}

    \centering
    
        \caption{\justifying \small Effect of SHI strength $K_s$ on output-phase binarization, mixed curvature, and hidden-layer response for different values of $\alpha$.
(a) Output-phase deviation from the binary manifold, $\epsilon_y$, as a function of $K_s$. The deviation decreases with increasing $K_s$, reaching numerical zero at progressively smaller values of $K_s$ as $\alpha$ approaches $1/2$. At $\alpha=1/2$, the output phases remain numerically binary throughout the investigated range.
(b) Mixed hidden--output curvature $\|H_{hy}\|$ versus $K_s$. The curvature depends strongly on $|2\alpha-1|$ but varies weakly with $K_s$, becoming numerically negligible at $\alpha=1/2$.
(c) Hidden-layer response $\|\delta\bm{u}\|$ versus $K_s$. The response decreases with increasing $K_s$ and becomes negligible as the output phases approach the binary manifold.}
    \label{nearbinary}
\end{figure*}

Increasing the SHI strength drives the oscillator phases toward the binary values $0$ and $\pi$, causing $\cos u_i$ and $\cos v_a$ to approach $\pm1$. As shown in Eq.~\eqref{eq:loss_force}, the output nudging force is proportional to $\sin v_a$ and therefore decreases as the output phases approach the binary manifold, even when the prediction differs from the target. The mixed hidden--output curvature behaves differently. From Eq.~\eqref{eq:mixed_hessian_binary}, its binary-limit value is proportional to $2\alpha-1$ and remains finite for nonzero couplings when $\alpha\neq1/2$. Thus, increasing SHI can weaken the output nudging force without producing a comparable reduction in the cross-layer curvature.

Figure~\ref{fig:sussweep} examines the influence of SHI strength $K_s$ on the hidden-layer response at fixed $\alpha=0$. Figure~\ref{fig:sussweep}(a) shows that the phase deviations $\epsilon_h$ and $\epsilon_y$ decrease with increasing $K_s$, confirming that stronger SHI drives both hidden and output oscillators closer to the binary manifold. As the output phases approach $0$ or $\pi$, the factor $\sin v_a$ in Eq.~\eqref{eq:loss_force} decreases, thereby reducing the output nudging force. Consistent with this behavior, Fig.~\ref{fig:sussweep}(b) shows that $\|\bm q_y\|$ decreases by approximately three orders of magnitude, whereas the output-to-hidden susceptibility $\|T_\alpha\|$ changes by only a factor of $\sim2.4$. Consequently, the hidden-layer response $\|\delta\bm u\|$ closely follows the decrease in the nudging force. Figure~\ref{fig:sussweep}(c) further shows that the mixed curvature $\|H_{hy}\|$ varies by less than $2\%$ over the investigated range of $K_s$. The remaining variation in $\|T_\alpha\|$ can therefore be attributed primarily to the inverse-curvature factors $H_{hh}^{-1}$ and $S_y^{-1}$. For a positive-definite Hessian, the largest eigenvalue of each inverse is the reciprocal of the smallest eigenvalue of the corresponding matrix. Thus, smaller eigenvalues indicate greater susceptibility to perturbations. In the present bipartite network, $H_{hh}$ is diagonal, so its smallest eigenvalue is given by $\min\operatorname{diag}(H_{hh})$.

As $K_s$ increases, $\min\operatorname{diag}(H_{hh})$ increases, indicating greater hidden-layer stiffness and a reduced response to perturbations. In contrast, $\lambda_{\min}(S_y)$ varies nonmonotonically, reaching a minimum near $K_s=0.7$, where the effective output curvature softens and the output susceptibility increases. These competing trends help explain why $\|T_\alpha\|$ exhibits a comparatively modest variation over the $K_s$ sweep, despite substantial changes in the individual curvature factors.

These observations suggest that SHI and the interpolation parameter $\alpha$ provide complementary means of controlling the hidden-layer response. In the near-binary regime, SHI primarily affects the nudging force through the output-phase deviations, whereas $\alpha$ directly controls the leading contribution to the mixed hidden--output curvature. 

Figure~\ref{nearbinary} examines their combined influence by extending the $K_s$ sweep to six values of $\alpha \in \{0, 0.1, 0.2,0.3,0.4,0.5\}$. Figure~\ref{nearbinary}(a) shows that the output-phase deviation $\epsilon_y$ decreases with increasing $K_s$. The onset of numerical binarization occurs at progressively smaller values of $K_s$ as $\alpha$ approaches $1/2$. At the midpoint, the measured output-phase deviations are negligible throughout the investigated range. In contrast, Fig.~\ref{nearbinary}(b) shows that $\|H_{hy}\|$ depends strongly on $\alpha$ but varies only weakly with $K_s$. Its magnitude follows the dependence on $|2\alpha-1|$ predicted by Eq.~\eqref{eq:mixed_hessian_binary} and becomes numerically negligible at $\alpha=1/2$.

\begin{figure*}
  \centering
         \includegraphics[width=1.01\linewidth]{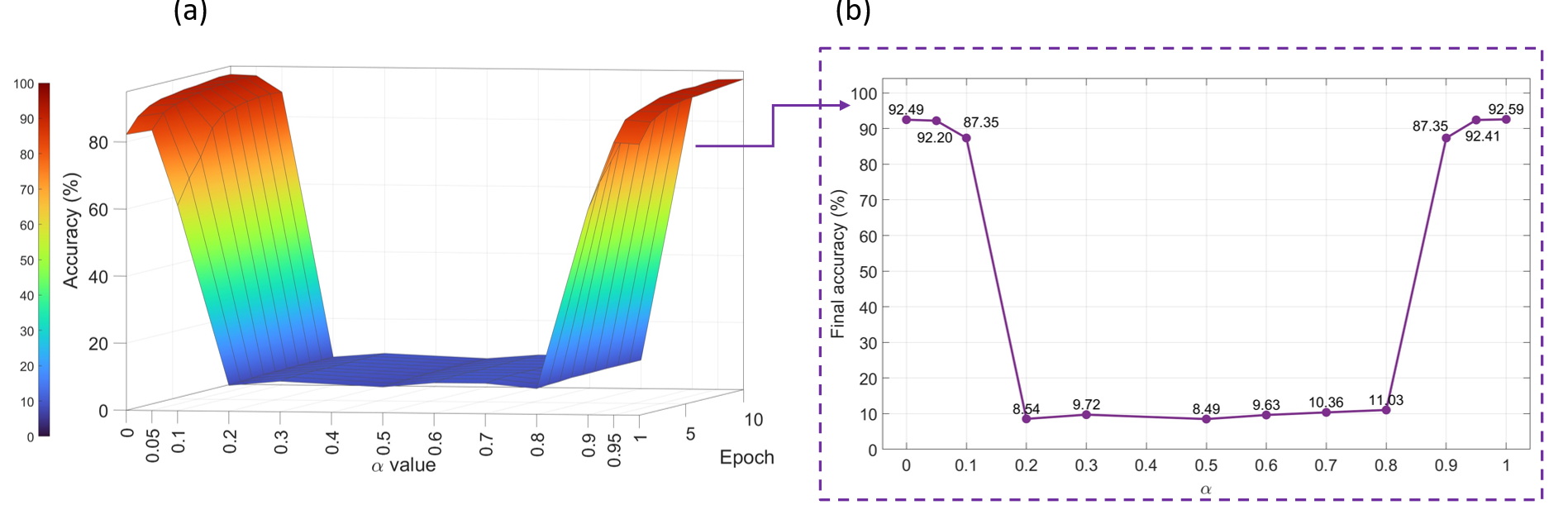}
  \caption{\justifying \small Test accuracy of the oscillator network trained using EP for different values of the coupling interpolation parameter $\alpha$ over 10 epochs. (a) Evolution of test accuracy with training epoch for different values of $\alpha$. (b) Final test accuracy as a function of $\alpha$ after 10 training epochs.}
  \label{accuracy}
\end{figure*}

The resulting hidden-layer response is shown in Fig.~\ref{nearbinary}(c). The response is largest at small $K_s$ and $\alpha=0$, where both the nudging force and cross-layer curvature are appreciable. Increasing $K_s$ suppresses the response by reducing the output nudging force, while moving $\alpha$ toward $1/2$ reduces the mixed curvature and further weakens perturbation transmission. At the midpoint, the hidden response becomes numerically negligible throughout the investigated range of $K_s$.

Together, these results demonstrate that the hidden-layer response depends on both the output nudging force and the cross-layer susceptibility. SHI primarily controls the former through phase binarization, while $\alpha$ controls the latter through the mixed curvature.

\subsection{Training accuracy assessment}
We next examine how the coupling parameter $\alpha$ influences the learning performance of the network. Classification accuracy is evaluated on the 10,000 MNIST test images, with the predicted class determined by the output oscillator having the largest $\cos v_a$ at the free equilibrium.

Figure~\ref{accuracy} shows the test accuracy as a function of $\alpha\in[0,1]$ over ten training epochs, with all other hyperparameters held fixed at $K_{s,h}=0.05$ and $K_{s,y}=0.07$. The highest accuracies are obtained at the pure phase-difference ($\alpha=0$) and phase-sum ($\alpha=1$) coupling limits. As $\alpha$ approaches $1/2$, the accuracy decreases substantially, reaching approximately chance level over a broad region surrounding the midpoint.

This behavior is consistent with the suppression of the hidden-layer response predicted by Eq.~\eqref{eq:du_scaling}. Near the midpoint, the mixed curvature decreases with $|2\alpha-1|$, weakening the transmission of output perturbations to the hidden layer. However, the scaling analysis alone does not explain the full width of the observed low-accuracy region. Equation~\eqref{eq:du_scaling} identifies the regime $|2\alpha-1|\lesssim\epsilon_h\epsilon_y$ in which the residual near-binary curvature becomes comparable to the leading $\alpha$-dependent contribution, but it does not directly determine learning performance.

An additional effect is evident in Fig.~\ref{nearbinary}(a), which shows that the output-phase deviation $\epsilon_y$ also decreases as $\alpha$ approaches $1/2$. Consequently, the output nudging force weakens while the cross-layer curvature is simultaneously suppressed. These two effects reduce the hidden-layer response over a potentially broader range of $\alpha$ than would be expected from the curvature cancellation alone. The observed accuracy valley is consistent with this combined suppression, although its quantitative width also depends on the training dynamics.

\section{Conclusion}

In this work, we investigated how the form of oscillator coupling influences equilibrium propagation in a bipartite oscillator Ising machine while preserving the same binary Ising Hamiltonian. We established an exact reflection symmetry relating systems at $\alpha$ and $1-\alpha$, extending to their nonlinear trajectories, equilibria, Hessian spectra, and EP parameter updates. At the symmetry point $\alpha=1/2$, the mixed hidden--output curvature vanishes on the binary manifold, suppressing the transmission of output perturbations to the hidden layer even though the forward interaction remains finite. Numerical simulations confirm the reflection symmetry and demonstrate the suppression of the hidden-layer response near the midpoint.

Our results establish that reproducing the desired Ising energy alone is insufficient to characterize the learning behavior of oscillator networks. The learning signal received by the hidden layer depends on both the nudging force applied at the output and the susceptibility governing its transmission across layers. We further showed that second-harmonic injection and coupling symmetry primarily influence these two factors through phase binarization and mixed curvature, respectively. These findings identify output-to-hidden susceptibility as an important design consideration for oscillator-based physical learning systems. While the present work focuses on equilibrium propagation, similar considerations may be relevant to other physical learning frameworks~\cite{munakata2004hebbian,altman2024experimental,lopez2023self}.

\section*{Data availability}
The MNIST dataset~\cite{lecun1998gradient} analyzed in this study is publicly
available. The numerical data underlying the figures are available from the corresponding author upon request.
\section*{Code availability}
The simulation code and trained models are available from the corresponding author upon request.

\begin{acknowledgments}
{ This material is based upon work supported in part by National Science Foundation grants (2328961)}
\end{acknowledgments}

\appendix






\vspace{1in}

\section{Relation between the EP estimator and the implicit equilibrium gradient}
\label{app:ep_gradient}

Here, we establish the relationship between the EP parameter updates and the gradient obtained by differentiating the training objective through the equilibrium state.

The free equilibrium $\bm z^{*(0)}$ satisfies
\begin{equation}
\nabla_{\bm z}E(\bm z^{*(0)};\vartheta)=0.
\end{equation}
Differentiating this equilibrium condition with respect to the trainable parameters $\vartheta$ gives
\begin{equation}
H\frac{\partial\bm z^{*(0)}}{\partial\vartheta}
+
\nabla^2_{\bm z\vartheta}E=0,
\end{equation}
where $H=\nabla^2_{\bm z\bm z}E$ is the Hessian evaluated at the free equilibrium. Assuming that $H$ is nonsingular, we obtain
\begin{equation}
\frac{\partial\bm z^{*(0)}}{\partial\vartheta}
=
-H^{-1}\nabla^2_{\bm z\vartheta}E.
\label{eq:equilibrium_derivative}
\end{equation}

The training objective is the output loss evaluated at the free equilibrium,
\begin{equation}
\mathcal L(\vartheta)
=
\ell\bigl(\bm z^{*(0)}(\vartheta),\hat{\bm y}\bigr).
\end{equation}
Since the loss has no explicit dependence on $\vartheta$, applying the chain rule and substituting Eq.~\eqref{eq:equilibrium_derivative} yields
\begin{equation}
\nabla_\vartheta\mathcal L
=
-\bigl(\nabla^2_{\bm z\vartheta}E\bigr)^{\mathsf T}
H^{-1}\nabla_{\bm z}\ell.
\label{eq:implicit_gradient}
\end{equation}
Using the linear response to output nudging,
\begin{equation}
\delta\bm z
=
-H^{-1}\nabla_{\bm z}\ell,
\end{equation}
the implicit gradient becomes
\begin{equation}
\nabla_\vartheta\mathcal L
=
\bigl(\nabla^2_{\bm z\vartheta}E\bigr)^{\mathsf T}
\delta\bm z.
\label{eq:implicit_gradient_response}
\end{equation}

We now compare this result with the EP gradient estimator. The one-sided EP estimator of the negative gradient is
\begin{equation}
\bm g_{\mathrm{EP}}^{+}
=
-\frac{1}{\beta}
\left[
\partial_\vartheta E(\bm z^{*(\beta)})
-
\partial_\vartheta E(\bm z^{*(0)})
\right].
\label{eq:ep_one_sided}
\end{equation}
Expanding $\partial_\vartheta E$ around the free equilibrium and using
$\bm z^{*(\beta)}
=
\bm z^{*(0)}+\beta\delta\bm z+\mathcal O(\beta^2)$
gives
\begin{equation}
\bm g_{\mathrm{EP}}^{+}
=
-\bigl(\nabla^2_{\bm z\vartheta}E\bigr)^{\mathsf T}
\delta\bm z
+
\mathcal O(\beta).
\end{equation}
Comparing with Eq.~\eqref{eq:implicit_gradient_response}, we obtain
\begin{equation}
\bm g_{\mathrm{EP}}^{+}
=
-\nabla_\vartheta\mathcal L
+
\mathcal O(\beta).
\label{eq:ep_gradient_equivalence}
\end{equation}

For symmetric EP, the estimator is constructed using equilibria obtained under positive and negative nudging:
\begin{equation}
\bm g_{\mathrm{sym}}
=
-\frac{1}{2\beta}
\left[
\partial_\vartheta E(\bm z^{*(\beta)})
-
\partial_\vartheta E(\bm z^{*(-\beta)})
\right].
\label{eq:ep_symmetric}
\end{equation}
Provided the equilibrium branch and energy derivatives are sufficiently smooth near $\beta=0$, the symmetric difference cancels the even-order terms in the Taylor expansion, yielding
\begin{equation}
\bm g_{\mathrm{sym}}
=
-\bigl(\nabla^2_{\bm z\vartheta}E\bigr)^{\mathsf T}
\delta\bm z
+
\mathcal O(\beta^2).
\end{equation}
Therefore,
\begin{equation}
\bm g_{\mathrm{sym}}
=
-\nabla_\vartheta\mathcal L
+
\mathcal O(\beta^2).
\label{eq:ep_symmetric_equivalence}
\end{equation}

Thus, both EP estimators converge to the negative implicit equilibrium gradient as $\beta\rightarrow0$, with the symmetric estimator reducing the leading finite-nudging error from $\mathcal O(\beta)$ to $\mathcal O(\beta^2)$.

\
\bibliographystyle{apsrev4-2}
\bibliography{refrences}

\end{document}